\documentclass[conference]{IEEEtran}
\usepackage{cite}
\usepackage{amsmath,amssymb,amsfonts}
\usepackage{graphicx}
\usepackage{booktabs}
\usepackage{xcolor}
\usepackage{xurl}
\usepackage{enumitem}
\usepackage{standalone}
\usepackage{tikz}
\usepackage{makecell}
\usetikzlibrary{positioning, arrows.meta, shapes.geometric, fit, backgrounds, calc, svg.path}
\definecolor{otblue}{HTML}{0072B2}
\definecolor{otorange}{HTML}{E69F00}
\definecolor{otteal}{HTML}{009E73}
\definecolor{otpurple}{HTML}{CC79A7}
\definecolor{otgray}{HTML}{5A5A5A}
\definecolor{brandnpm}{HTML}{CB3837}
\definecolor{brandpypi}{HTML}{3775A9}
\definecolor{brandrust}{HTML}{000000}
\definecolor{branddepsdev}{HTML}{32E0C4}
\definecolor{branddepsdevdark}{HTML}{1AAC94}
\definecolor{brandgithub}{HTML}{181717}

\newcommand{\para}[1]{\par\smallskip\noindent\textbf{#1.}\ \ignorespaces}
\newcommand{\stepnum}[1]{\tikz[baseline=(c.base)]\node[circle, fill=otgray!90!black, text=white, font=\sffamily\tiny\bfseries, inner sep=0pt, minimum size=0.32cm] (c) {#1};}

\newcommand{\tool}{\textsc{mttx}}
\newcommand{\opt}[1]{\textsl{\small{#1}}}
\newcommand{\repofile}[1]{\href{https://github.com/imranur-rahman/dependency-update-metrics/blob/main/#1}{\texttt{#1}}}
\usepackage[most]{tcolorbox}

\usepackage[colorlinks=true, urlcolor=otblue, linkcolor=black, citecolor=black]{hyperref}

\begin{document}
\bstctlcite{IEEEexample:BSTcontrol}

\title{Beyond Stars: \tool{} Brings Developer Responsiveness to Dependency Decisions, \\from Dashboard to IDE to GitHub Actions}

\author{\IEEEauthorblockN{Imranur Rahman\IEEEauthorrefmark{1}, Ranindya Paramitha\IEEEauthorrefmark{2}, Mahzabin Tamanna\IEEEauthorrefmark{1}, Laurie Williams\IEEEauthorrefmark{1}}
\IEEEauthorblockA{\IEEEauthorrefmark{1}North Carolina State University, \{irahman3, mtamann, lawilli3\}@ncsu.edu\\
\IEEEauthorrefmark{2}University of Bristol, ranindya.paramitha@bristol.ac.uk}}

\maketitle

\begin{abstract}
Modern software reuses open-source packages as dependencies, and developers often choose among candidate packages with cheap signals, such as GitHub stars and download counts.
However, such signals are easy to fake, so developers need signals that represent ``observed behavior,'' such as how quickly the developers of a package respond to new dependency releases and to security fixes.
Prior work computed these responsiveness metrics, MTTU (Mean Time To Update) dependencies and MTTR (Mean Time To Remediate) vulnerable dependencies, but only with a research prototype for a fixed set of packages.
Therefore, developers cannot easily benefit from them or integrate them into their workflow.
We present \tool{}, an open-source tool that computes MTTU and MTTR with accurate historical dependency resolution for any package and given time window, and brings them into differnet phases of developer workflow.
\tool{} resolves each version constraint of npm, PyPI, and Cargo packages as of a past date, with data from the native registries or deps.dev and vulnerability data from OSV.
We validated \tool{} with 328 automated tests, and for 90 packages from prior work, we manually verified its dependency resolution at each historical release point and its MTTU and MTTR computation.
Developers can use \tool{} in an interactive web dashboard, an IDE extension, and a GitHub Action, which are all available from the project homepage at \url{https://imranur-rahman.github.io/dependency-update-metrics/}.
We envision that package registries (e.g., npm) and metadata aggregators and intelligence platforms (e.g., deps.dev) will display these values on the package page, alongside download counts and stars, to make them easier for developers to access.
A demo video is at \url{https://youtu.be/MZqjDnJ8bEQ}.
\end{abstract}


\section{Introduction}
\label{sec:intro}

\noindent Modern software reuses open-source packages as dependencies, because software reuse speeds up development and lowers cost~\cite{mohagheghi2007reuse}.
However, a project that uses a vulnerable version of a dependency can also inherit its known vulnerabilities~\cite{decanImpactSecurityVulnerabilities2018}, so the choice of a dependency is also a security decision.
To choose among candidate packages, developers use signals such as GitHub stars, download counts, and recent activity~\cite{lariosvargasSelectingThirdpartyLibraries2020}.

\noindent However, GitHub stars and download counts are cheap signals and so are easy to fake.
For example, He et al. found about six million suspected fake stars on GitHub~\cite{heFakeStars2026}, and a review of 252 web sources and 870 Reddit threads found that faking such signals costs less than earning them, which turns the software supply chain into a ``market for lemons''~\cite{paramithaLemons2026}.
To avoid such a market for dependencies, developers need signals that come from the ``observed behavior'' of maintainers.

\noindent One such signal is the responsiveness of the developers of a package: how quickly they keep their own dependencies up to date and adopt upstream security fixes~\cite{chinthanetLagsReleaseAdoption2021}.
This signal reflects what the developers did (``observed behavior'') since developers' responsiveness is measured from the release history of the package over time.
Unlike technical lag~\cite{gonzalez-barahonaTechnicalLagSoftware2017}, which measures how far a package lags at one moment, MTTU (Mean Time To Update) and MTTR (Mean Time To Remediate)~\cite{rahman2026howquickly} measure how quickly developers \emph{respond} to new dependency releases and to published security fixes over time.
A package with lower MTTU and MTTR keeps its dependencies current and patched, so between two candidates, a developer can prefer the one with lower values if developers care about security.

\noindent However, prior work computed MTTU and MTTR with a research prototype for a fixed set of packages~\cite{rahman2026howquickly}, and so these values are not readily available where they decide on dependencies: when they select a package, while they code, and in continuous integration (CI).
Consequently, developers cannot easily benefit from these metrics or integrate them into their workflow.
A reusable tool that provides these metrics is hard to build to make it accessible to various stakeholders (e.g., developers, researchers, security teams, and tool builders) since the tool must be accurate, verifiable, configurable, easy to use, and available for all stakeholders.
To our knowledge, no existing tool computes MTTU and MTTR for any package and time window and brings them into the developer workflow as well as other stakeholders (\S\ref{sec:related}).

\noindent \textit{The goal of \tool{} is to give developers a signal that comes from the observed behavior of the developers of a package, as recorded in its release history, at different phases of the software development lifecycle and for various stakeholders.}
Developers can use \tool{} to compare candidate packages, researchers and security teams to analyze dependencies at scale, and builders of security-metric tools, such as the OpenSSF Scorecard~\cite{zahanOpenSSFScorecardPath2023}, to add a signal of maintenance.

\noindent \tool{} implements the MTTU and MTTR algorithm~\cite{rahman2026howquickly} for npm, PyPI, and Cargo packages, with data from the native registries or deps.dev and vulnerability data from OSV (\S\ref{sec:architecture}).
In addition, three front ends, an interactive web dashboard, an IDE extension, and a GitHub Action, run this one CLI and make it accessible to various stakeholders (\S\ref{sec:features}).
In a manual validation of 90 packages from prior work~\cite{rahman2026howquickly}, we verified the dependency resolution of \tool{} at each historical release point and its MTTU and MTTR computation.

\noindent We envision that package registries, such as npm, and metadata platforms, such as deps.dev, will show MTTU and MTTR on the package page, next to download counts and stars, where developers choose their dependencies.
Until then, developers can use \tool{} in an interactive web dashboard to select, an IDE extension to code and develop, and a GitHub Action in CI (\S\ref{sec:availability}).
\noindent This paper makes three contributions:
\begin{itemize}[leftmargin=*,nosep]
\item \tool{}, an open-source tool that computes MTTU and MTTR and integrates them into different phases of developer workflow (\S\S\ref{sec:features}, \ref{sec:architecture}, and~\ref{sec:availability});
\item A validation of \tool{} with 328 automated tests and a manual validation of 90 packages (\S\ref{sec:validation}); and
\item Four guidelines, from the lessons that we learned, for future builders of software engineering research tools (\S\ref{sec:lessons}).
\end{itemize}

\section{Features}
\label{sec:features}
\noindent \tool{} brings MTTU and MTTR to different phases of the software development lifecycle for better dependency management and for various stakeholders.
On a high level, \tool{} CLI runs the core computation of \tool{}, and three front ends utilize the CLI in different development phases (Fig.~\ref{fig:architecture}).

\para{Select: interactive web dashboard}
Developers compare the MTTU and MTTR of up to five candidate packages from one ecosystem, without an installation, through our interactive web dashboard.
The dashboard shows a bar chart (and, at every release, line charts), a table, and a layman's term explanation of each value along with its limitations.

\para{Develop: IDE extension}
In VS Code and in the editors that use Open VSX, such as Cursor, Windsurf, VSCodium, the extension shows the MTTU and MTTR of each dependency as an inline hint in \texttt{package.json} for npm, \texttt{pyproject.toml} for PyPI, and \texttt{Cargo.toml} for crates.io packages, and a separate command compares the dependency with alternatives.

\para{Integrate: GitHub Action}
The Action has two modes.
In the \emph{packages} mode, the Action reports the MTTU and MTTR of chosen packages in CI on a specified interval, for example each week, as step outputs, a job summary, and badges.
In the \emph{pull-request} mode, the Action comments on a pull request with the values of the dependencies that the pull request adds and a brief explanation of the values.
In both modes, the Action fails when the computed MTTX value exceeds a limit that the team set in the Action's configuration.

\para{Analyze and reproduce: CLI}
Users can analyze one package, every past release of a package (\opt{per-release}), or a CSV file of packages in parallel (\opt{input-csv}), for example a whole ecosystem.
Optionally they can choose the data source (the native registries or deps.dev~\cite{depsdev}), and a recency weight.
\opt{resume}, \opt{timeouts}, and \opt{memory limits} keep long batches running.

\section{Tool Architecture}
\label{sec:architecture}
\begin{figure}[t]
\centering
\def\logoh{0.5cm}%
\def\boxh{0.9}%
\def\ytop{3.55}%
\def\ymid{2.1}%
\def\yusers{-2.05}%
\def\xin{0}%
\def\xcli{0}%
\def\xres{3.4}%
\def\xana{7.35}%
\def\inpkglabel{Package(s)}%
\def\inwinlabel{Time window}%
\def\inconflabel{Configurations}%
\def\clilabel{CLI}%
\def\registrylabel{Package registries}%
\def\cachelabel{Resolver Cache}%
\def\resolverlabel{Historical\\Dependency Resolver}%
\def\indexlabel{OSV Index}%
\def\analyzerlabel{Dependency\\Analyzer}%
\def\reportlabel{{\hdrfont Reporting}\\[-1pt]{\tiny MTTU, MTTR}}%
\def\newecolabel{+ new ecosystem}%
\def\metadatalabel{package metadata}%
\def\advisorylabel{advisory data}%
\def\timelinelabel{dependency\\timelines}%
\def\dashlabel{Interactive\\Dashboard}%
\def\idelabel{IDE extensions}%
\def\actionlabel{GitHub Action}%
\def\runlabel{runs}%
\def\resultlabel{results}%
\graphicspath{{icons/}{figures/icons/}}%
\InputIfFileExists{figures/logos}{}{\input{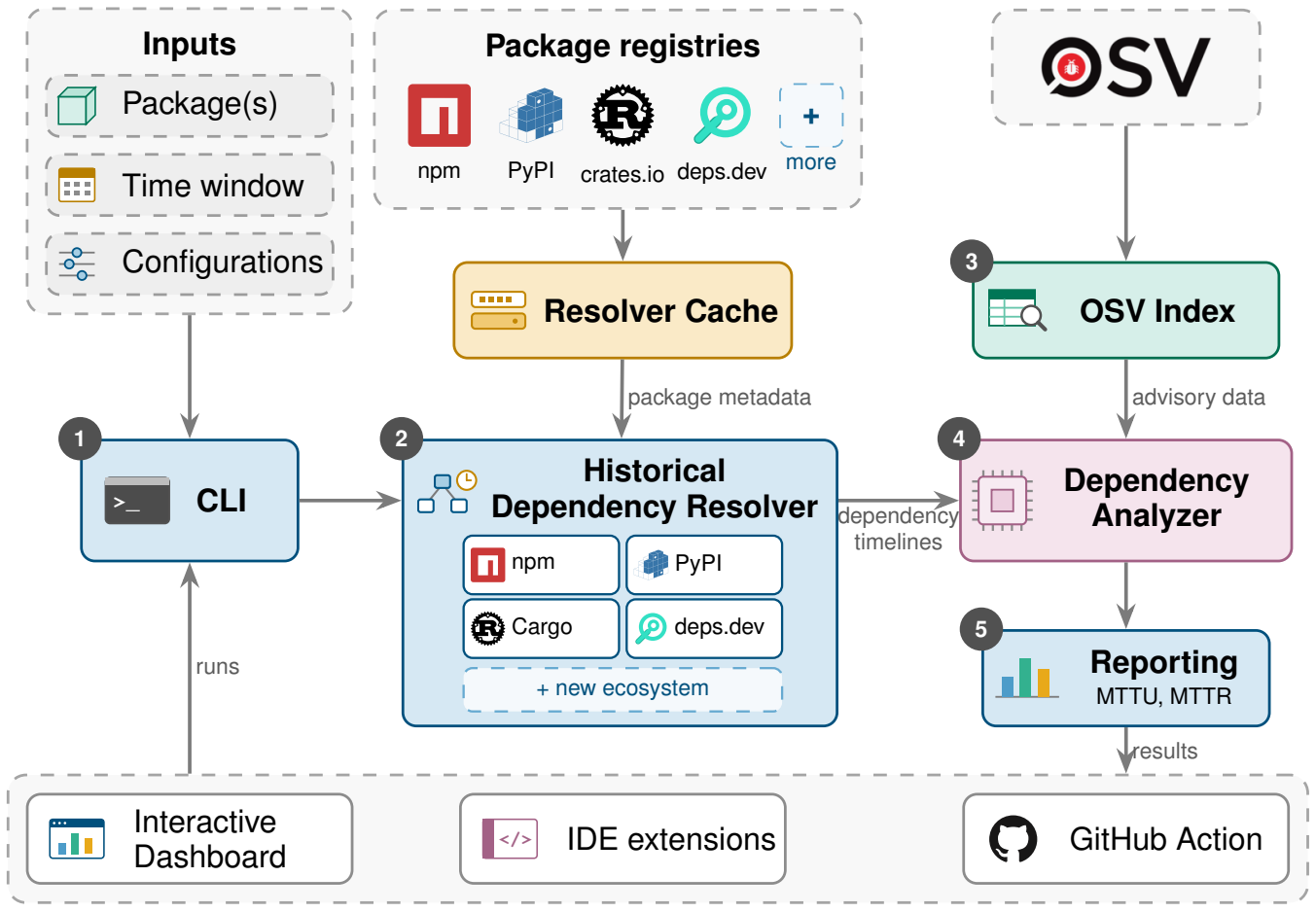}}%
\providecommand{\hdrfont}{\fontfamily{phv}\fontseries{b}\selectfont}%
\begin{tikzpicture}[
    >={Stealth[length=2mm]},
    comp/.style={
      rounded corners=3pt, line width=0.6pt, align=center,
      minimum height=\boxh cm, font=\sffamily\scriptsize, inner sep=2pt,
    },
    external/.style={comp, draw=otgray!60, fill=otgray!10, dashed},
    input/.style={external, minimum width=2.25cm, minimum height=0.48cm},
    process/.style={comp, minimum width=1.9cm, draw=otblue!70!black, fill=otblue!16},
    core/.style={comp, minimum width=2.9cm, draw=otpurple!80!black, fill=otpurple!20},
    slot/.style={rounded corners=2pt, line width=0.5pt, align=center,
      draw=otblue!70!black, fill=white, font=\sffamily\tiny,
      minimum width=1.12cm, minimum height=0.34cm, inner sep=1pt},
    newslot/.style={slot, draw=otblue!60, dashed, fill=otblue!4, text=otblue!70!black},
    store/.style={
      shape=cylinder, shape border rotate=90, aspect=0.22,
      draw=otteal!70!black, fill=otteal!16, line width=0.6pt,
      minimum width=2.1cm, minimum height=0.8cm, align=center,
      font=\sffamily\scriptsize, inner sep=1pt,
    },
    cachestore/.style={store, draw=otorange!80!black, fill=otorange!22},
    link/.style={draw=otgray!80, line width=0.8pt, ->},
    group/.style={draw=otgray!55, dashed, rounded corners=4pt, line width=0.6pt,
      inner sep=4pt, fill=otgray!4},
    resgroup/.style={draw=otblue!70!black, rounded corners=3pt, line width=0.6pt,
      inner sep=3pt, fill=otblue!16},
    grouptitle/.style={font=\scriptsize\hdrfont, inner sep=1pt, align=center},
    edgelabel/.style={font=\sffamily\tiny, text=otgray, align=left, inner sep=1pt},
    logolabel/.style={font=\sffamily\tiny, inner sep=0.5pt},
    stepno/.style={circle, fill=otgray!90!black, text=white, font=\sffamily\tiny\bfseries,
      inner sep=0pt, minimum size=0.34cm},
  ]

  \coordinate (colin) at (\xin, 0);
  \coordinate (colana) at (\xana, 0);

  \foreach \n/\dx/\dy/\lg/\lab in {snpm/-0.64/0.1/\npmlogo/npm, spypi/0.64/0.1/\pypilogo/PyPI,
                                   scargo/-0.64/-0.4/\rustlogo/Cargo, sdeps/0.64/-0.4/\depsdevlogo/deps.dev}{
    \node[slot, minimum width=1.22cm, minimum height=0.46cm] (\n) at ($(\xres, 0) + (\dx, \dy)$) {};
    \node[inner sep=0pt] at ([xshift=0.2cm]\n.west) {\scalebox{0.55}{\usebox{\lg}}};
    \node[font=\sffamily\tiny, anchor=west, inner sep=0pt] at ([xshift=0.38cm]\n.west) {\lab}; }
  \node[newslot, minimum width=2.5cm] (snew) at ($(\xres, 0) + (0, -0.88)$) {\newecolabel};
  \node[grouptitle, above=0.04cm of $(snpm.north)!0.5!(spypi.north)$, xshift=0.25cm] (restitle) {\resolverlabel};
  \node[left=0.06cm of restitle, inner sep=0pt] (resicon) {\usebox{\deptreeicon}};
  \begin{scope}[on background layer]
    \node[resgroup, fit=(restitle)(resicon)(snpm)(spypi)(scargo)(sdeps)(snew)] (resolver) {};
  \end{scope}

  \node[core, minimum width=2.6cm, minimum height=0.95cm, anchor=north] (analyzer) at (colana |- resolver.north) {};
  \node[inner sep=0pt] at ([xshift=0.34cm]analyzer.west) {\usebox{\cpuicon}};
  \node[font=\scriptsize\hdrfont, align=center] at ([xshift=0.24cm]analyzer.center) {\analyzerlabel};
  \node[process, minimum width=2.25cm, minimum height=0.75cm, anchor=south] (report) at (colana |- resolver.south) {};
  \node[inner sep=0pt] at ([xshift=0.36cm]report.west) {\usebox{\charticon}};
  \node[font=\sffamily\scriptsize, align=center] at ([xshift=0.3cm]report.center) {\reportlabel};

  \node[process, minimum width=1.7cm, minimum height=0.95cm, anchor=north] (cli) at (colin |- resolver.north) {};
  \node[inner sep=0pt] at ([xshift=0.45cm]cli.west) {\usebox{\termicon}};
  \node[font=\scriptsize\hdrfont] at ([xshift=0.25cm]cli.center) {\clilabel};

  \node[comp, minimum width=2.65cm, minimum height=0.75cm, draw=otorange!80!black, fill=otorange!22] (cache) at (\xres, \ymid) {};
  \node[inner sep=0pt] at ([xshift=0.36cm]cache.west) {\usebox{\cacheicon}};
  \node[font=\scriptsize\hdrfont] at ([xshift=0.3cm]cache.center) {\cachelabel};
  \node[comp, minimum width=2.4cm, minimum height=0.75cm, draw=otteal!70!black, fill=otteal!16] (index) at (\xana, \ymid) {};
  \node[inner sep=0pt] at ([xshift=0.36cm]index.west) {\usebox{\indexicon}};
  \node[font=\scriptsize\hdrfont] at ([xshift=0.25cm]index.center) {\indexlabel};

  \node (lnpm)   at ($(\xres, \ytop) + (-1.44, 0.08)$) {\usebox{\npmlogo}};
  \node (lpypi)  at ($(\xres, \ytop) + (-0.72, 0.08)$) {\usebox{\pypilogo}};
  \node (lrust)  at ($(\xres, \ytop) + ( 0.0, 0.08)$) {\usebox{\rustlogo}};
  \node (ldeps) at ($(\xres, \ytop) + (0.78, 0.08)$) {\usebox{\depsdevlogo}};
  \node[draw=otblue!60, dashed, rounded corners=2pt, fill=otblue!4, text=otblue!70!black,
    font=\sffamily\scriptsize\bfseries, minimum size=14pt, inner sep=0pt] (lnew)
    at ($(\xres, \ytop) + (1.48, 0.08)$) {+};
  \node[logolabel, text=otblue!70!black, below=0.05cm of lnew] {more};
  \node[logolabel, below=0.0cm of ldeps] {deps.dev};
  \node[logolabel, below=0.0cm of lnpm]  {npm};
  \node[logolabel, below=0.0cm of lpypi] {PyPI};
  \node[logolabel, below=0.0cm of lrust] {crates.io};
  \node[grouptitle, above=0.0cm of lrust.north] (regtitle) {\registrylabel};
  \coordinate (regbottom) at ($(lnpm.south)+(0,-0.22)$);
  \begin{scope}[on background layer]
    \node[group, fit=(regtitle)(lnpm)(lnew)(regbottom)] (registries) {};
  \end{scope}

  \node[group, minimum width=2.1cm, minimum height=0.9cm, anchor=north] (osv) at (colana |- registries.north) {};
  \node (losv) at (osv.center) {\usebox{\osvlogo}};

  \node[grouptitle, anchor=north] (intitle) at ([yshift=-4pt]colin |- registries.north) {Inputs};
  \node[input, below=0.06cm of intitle] (inpkg) {};
  \node[input, below=0.12cm of inpkg] (inwin) {};
  \node[input, below=0.12cm of inwin] (inconf) {};
  \foreach \n/\ic/\lab in {inpkg/\pkgicon/\inpkglabel, inwin/\calicon/\inwinlabel, inconf/\sliicon/\inconflabel}{
    \node[inner sep=0pt] at ([xshift=0.25cm]\n.west) {\usebox{\ic}};
    \node[font=\sffamily\scriptsize, anchor=west] at ([xshift=0.48cm]\n.west) {\lab}; }
  \begin{scope}[on background layer]
    \node[group, fit=(intitle)(inpkg)(inwin)(inconf)] (inputs) {};
  \end{scope}

  \foreach \n/\x/\ic/\lab in {udash/\xin/\dashicon/\dashlabel, uide/\xres/\ideicon/\idelabel,
                               uaction/\xana/\githubicon/\actionlabel}{
    \node[comp, minimum width=2.55cm, minimum height=0.7cm, draw=otgray!70, fill=white] (\n) at (\x, \yusers) {};
    \node[inner sep=0pt] at ([xshift=0.4cm]\n.west) {\scalebox{0.85}{\usebox{\ic}}};
    \node[font=\sffamily\scriptsize, align=left, anchor=west] at ([xshift=0.72cm]\n.west) {\lab}; }
  \begin{scope}[on background layer]
    \node[group, fit=(udash)(uide)(uaction)] (users) {};
  \end{scope}

  \draw[link] (inputs.south -| cli) -- (cli);
  \draw[link] (cli) -- (cli -| resolver.west);
  \draw[link] (resolver.east |- analyzer) -- node[edgelabel, below, align=center] {\timelinelabel} (analyzer);
  \draw[link] (analyzer) -- (report);
  \draw[link] (registries.south -| cache) -- (cache);
  \draw[link] (cache) -- node[edgelabel, right] {\metadatalabel} (resolver);
  \draw[link] (osv.south -| index) -- (index);
  \draw[link] (index) -- node[edgelabel, right] {\advisorylabel} (analyzer);
  \draw[link] (users.north -| cli) -- node[edgelabel, right] {\runlabel} (cli);
  \draw[link] (report) -- node[edgelabel, right] {\resultlabel} (users.north -| report);

  \node[stepno] at (cli.north west)      {1};
  \node[stepno] at (resolver.north west) {2};
  \node[stepno] at (index.north west)    {3};
  \node[stepno] at (analyzer.north west) {4};
  \node[stepno] at (report.north west)   {5};

\end{tikzpicture}
\caption{High level architecture overview of \tool{}.}
\label{fig:architecture}
\end{figure}

\noindent Fig.~\ref{fig:architecture} shows the five components of \tool{}, and at the bottom, the front ends of the CLI, which run it and provide the results to different phases of the software development lifecycle.

\para{Inputs and outputs}
The CLI (command-line interface) takes three inputs: the \emph{package(s)} (an ecosystem and a package name, or a CSV file of packages), the \emph{time window} $[t_s, t_e]$, and the \emph{configurations}.
The configurations turn on the features of \S\ref{sec:features}, such as the mode, the recency weight, and the data source.
In return, \tool{} outputs MTTU and MTTR for each package or release, as JSON or CSV.

\para{\stepnum{1} CLI}
The CLI first takes the inputs, builds the index of OSV~\cite{osv} vulnerability data if needed, and picks a resolver for each ecosystem.
In a batch run, \tool{} then groups the rows by package, so that all rows of a package share one fetch and one worker process.

\para{\stepnum{2} Historical Dependency Resolver}
Following the MTTU and MTTR definitions of Rahman et al.~\cite{rahman2026howquickly} (full algorithm in \repofile{ALGORITHM.md}), \tool{} splits the time window of each dependency into intervals at every new release of the package or of the dependency.
For each interval, the resolver then resolves two versions: (a) the version that the package's constraint selected at that time, and (b) the newest stable version available then.
To ensure correct historical dependency resolution, \tool{} uses only data published before the interval starts.
For npm and Cargo, \tool{} therefore filters the releases by publish time, much like npm's \texttt{--before} flag~\cite{npmbefore}, and then resolves the dependency.
\texttt{pip} has no such option, so we forked \texttt{pip} to add it for PyPI.
``Resolver Cache'' stores the registry data, because \tool{} may read each package many times.
Every resolver implements one interface (``PackageResolver''), so a new ecosystem needs only a new resolver implementation.

\para{\stepnum{3} OSV Index}
The OSV index stores the vulnerability data that the analyzer needs.
\tool{} turns the OSV data into one table, in which each row is the range of versions that one vulnerability affects, up to the first fixed version, with its severity.
\tool{} stores the table in memory, so that each lookup needs no API call.

\para{\stepnum{4} Dependency Analyzer}
The analyzer augments the intervals that the resolver \stepnum{2} produces.
An interval counts as \emph{outdated} if the selected version is not the latest one.
Similarly, it counts as \emph{unremediated} if the selected version has a known vulnerability whose fix was already published when the interval started.
Following Rahman et al.~\cite{rahman2026howquickly}, the analyzer then adds up, for each dependency $d$, the days in its outdated intervals $O_d$, and MTTU is the mean over the set $D$ of direct runtime dependencies:
\begin{equation}
\mathrm{TTU}_d = \sum_{i \in O_d} \Delta_i, \qquad \mathrm{MTTU} = \frac{1}{|D|} \sum_{d \in D} \mathrm{TTU}_d,
\label{eq:mttu}
\end{equation}
where $\Delta_i$ is the length of interval $i$ in days.
MTTR follows Eq.~\eqref{eq:mttu} with the unremediated intervals $U_d$ in place of $O_d$.
An optional recency weight $w$ favors recent intervals, because recent behavior may matter more to a user.
With a weight, $\mathrm{TTU}_d$ becomes the weighted mean $\sum_{i \in O_d} w_i \Delta_i / \sum_{i \in O_d} w_i$, where $w_i = w(t_e - t_i)$ depends on the age of interval $i$ at the end $t_e$ of the time window.

\para{\stepnum{5} Reporting}
Reporting writes a JSON file for one package, or one CSV row per package as soon as its analysis finishes.
Because each row is written at once, \opt{resume} can use these rows to skip finished work after a crash.
Optional worksheets list every interval, so that a reader can verify each value manually.

\section{Validation}
\label{sec:validation}
\noindent We validated \tool{} in two ways: automated tests check each part of \tool{}, and a manual validation with 90 packages.

\para{Automated tests}
We designed the tests after Kanewala and Bieman's review of testing scientific software~\cite{kanewala2014testing}.
Our 239 unit tests check each component in isolation: each resolver, the cache, the vulnerability index, and the reporting module.
Our 82 integration tests then check that the components work together, as they run the analyzer and the CLI over stubbed resolvers.
Among these tests, 40 regression tests reproduce the nine bugs that we found in the manual validation, so that a fixed bug cannot come back.
Finally, our 7 metamorphic tests change an input in a known way and check how the output must change~\cite{chen2018metamorphic}; for example, shifting all dates by the same amount must leave MTTU unchanged.
To check that these tests catch real faults, we injected two faults into the vulnerability index, and a metamorphic test and a test with a hand-computed answer each failed.
All 328 tests run offline on fixture data, so they are fast and repeatable, and CI runs them on three operating systems and three Python versions.

\para{Manual validation}
We used a manual validation to check \tool{} on packages, because tests with fixture data cannot show that \tool{} handles registry data correctly.
We randomly selected 90 packages (35 from npm, 35 from PyPI, and 20 from Cargo) from Rahman et al.~\cite{rahman2026howquickly} for manual validation.

\noindent We ran \tool{} on each package over its full history up to July~6, 2026, with deps.dev data and the inverse recency weight.
For each of the 90 packages, we manually verified the dependency resolution at each historical release point against the registry data of that date, and then recomputed MTTU and MTTR from the worksheets.
The first author performed the manual verification. 
Since the procedure involved deterministic version comparisons and arithmetic calculations, we documented each expected and observed value in the GitHub repository to make the verification reproducible.
The first round of manual verification found nine bugs, such as a stale version-list cache.
We fixed each bug and wrote regression tests that reproduce it on fixture data, with the inputs that triggered it and the values that we computed manually, so that a similar fault in the future fails a test.
We then repeated the manual verification for all 90 packages and found no errors or disagreements between the manually determined and tool-produced values.

\section{Lessons Learned}
\label{sec:lessons}
\noindent We built \tool{} iteratively, and in the iterations we faced four software engineering challenges.
We overcame each challenge with a guideline from prior software engineering research.
We report these lessons briefly, so that future builders of research tools for developers can avoid the same pitfalls.

\para{Metric Validity}
A metric should meet the validation criteria that fit its intended use~\cite{meneely2012validating}.
Our \tool{} counts only direct runtime dependencies, and its worksheets make every step of \tool{} open to verification.

\para{Temporal Consistency}
A value or computed metric about the past is correct only if it uses the data that existed at that time~\cite{gonzalez-barahonaTechnicalLagSoftware2017}.
So, for each interval, the resolver uses only the releases that were published before the interval starts.
Our resolver cache follows the same rule: the data of one released version never changes, so the cache stores it, but the list of versions of a package grows with each release, so \tool{} fetches the list again when a run asks for a later end date.
An early version of \tool{} reused the old list, and its reruns missed new releases which we found with manual verification.

\para{Fault Tolerance and Resource Utilization}
Software should keep working despite faults and use resources efficiently~\cite{iso25010}.
In \tool{}, \opt{resume} restarts a batch run from previous checkpoint to handle interruptions, and one process analyzes one package with a timeout and a memory limit.

\para{Modularity and Configurability}
Each module should encapsulate one of the ``design decisions which are likely to change''~\cite{parnas1972criteria} and a designed metric for developers should be configurable~\cite{meneely2012validating}.
In \tool{}, one resolver interface hides the ecosystems and the data sources, and the CLI options let a user choose the data source, analysis time window, and optional weighting mechanisms.

\section{Related Work}
\label{sec:related}
\para{Our prior work}
We defined MTTU and MTTR and the algorithm that \tool{} implements, evaluated the metrics ecosystem-wide, and analyzed if MTTU can be a proxy for MTTR when vulnerability data is scarce~\cite{rahman2026howquickly}.
In another study, we also compared pinning and floating version constraints for outdated and vulnerable dependencies~\cite{Rahman2025ASE}.
In a recent study, we open sourced \verb|Chrono-Resolution|~\cite{rahman2026chrono} dataset which contains historical dependency resolutions at release points for npm, PyPI, and crates.io packages but the dataset is generic and can be used for any dynamic metrics computation (e.g., dependency freshness, dependency update rhythm, etc.) and not limited to MTTU and MTTR.
Moreover, in similar other studies, we provided developers visibility into dependencies with security-sensitive API calls~\cite{rahmanWhatsPackageGetting2025}, and outlined research directions in software supply chain security~\cite{williamsResearchDirectionsSoftware2025}.

\para{Tools in the developer workflow}
Developers can see dependency signals in their workflow: deps.dev~\cite{depsdev} and the OpenSSF Scorecard~\cite{zahanOpenSSFScorecardPath2023} can show the dependencies, advisories, and security practices of a package, and bots such as Dependabot~\cite{alfadelUseDependabotSecurity2021} can update the dependencies of the developer's own project.
These tools show the state of a package at one moment or act on the developer's behalf, but none of them shows upstream developers' responsiveness or how quickly the developers of a candidate package responded over time.

\para{Novelty}
Our prior studies computed the metrics~\cite{rahman2026howquickly} and the dataset~\cite{rahman2026chrono} with research prototypes for a fixed set of packages, and provided developers no way to use the values in different phases of development workflow.
\tool{} turns the algorithm into a tested tool and integrates it into different phases of the developer workflow: an interactive web dashboard for dependency selection, an IDE extension for coding and development, and a GitHub Action for CI.

\section{Threats to Validity}
\label{sec:threats}
\para{Construct validity}
\tool{} measures direct dependencies only and skips pre-release versions, unless a constraint names one.
Registries do not record when a version was unpublished, so \tool{} cannot restore every version that was visible at a past date.
In addition, MTTR may underestimate the exposure to vulnerabilities, because the OSV index keeps only the ranges that have a fixed version.
As a result, a vulnerability that was never fixed does not count against MTTR.
We decided to exclude such cases, because a developer cannot update to a fix that does not exist.
In addition, a maintainer can lower MTTU and MTTR to game our metrics, for example by dropping dependencies or convincing the upstream maintainers not to fix vulnerabilities, both of which we consider extreme cases.
Nevertheless, developers should use our metrics together with other signals.

\para{Internal validity}
The values of \tool{} depend on its data sources, because deps.dev and OSV keep updating their records.
For example, native registries and deps.dev can resolve the same constraint differently, because deps.dev resolves old PyPI releases with a modern resolver~\cite{rahman2026chrono}.
A study should thus name the data source and the date of its OSV snapshot, so that others can repeat the run.

\para{External validity}
\tool{} covers npm, PyPI, and Cargo, and our manual validation used 90 packages from these three ecosystems.
As a result, our findings may not hold for other ecosystems, whose version rules differ.
Future work can add ecosystems that deps.dev already covers, such as Maven and Go, and can compute dependency freshness and update rhythm from the dependency timelines of \tool{}.
We will also propose to package registries that they show these values on their package pages.

\section{Availability}
\label{sec:availability}
\tool{} is open source under the MIT license, and its project homepage, \url{https://imranur-rahman.github.io/dependency-update-metrics/}, and the source repo is at \url{https://github.com/imranur-rahman/dependency-update-metrics}.
The CLI is on \href{https://pypi.org/project/dependency-update-metrics/}{PyPI} (\texttt{\small pip install dependency-update-metrics}), and its container image is \href{https://github.com/imranur-rahman/dependency-update-metrics/pkgs/container/mttx}{\texttt{ghcr.io/imranur-rahman/mttx}} on GHCR.
The interactive web dashboard runs on the project homepage, the IDE extension is on the \href{https://marketplace.visualstudio.com/items?itemName=imranur-rahman.mttx}{Visual Studio Marketplace} and \href{https://open-vsx.org/extension/imranur-rahman/mttx}{Open VSX}, and the GitHub Action is on the \href{https://github.com/marketplace/actions/mttx-dependency-update-metrics}{GitHub Marketplace}.
The demo video is at \url{https://youtu.be/MZqjDnJ8bEQ}.
The source code of \tool{}, as archived for this paper, is on Zenodo with the DOI \href{https://doi.org/10.5281/zenodo.23608337}{10.5281/zenodo.23608337}.

\bibliographystyle{IEEEtran}
\bibliography{references,zotero}

\end{document}